\documentclass[twocolumn,showpacs,preprintnumbers,superscriptaddress,amsmath,amssymb,prl, aps]{revtex4-2}

\usepackage{graphicx}

\usepackage[dvipsnames]{xcolor}
\usepackage[colorlinks=true,urlcolor=Blue,linkcolor=Blue,citecolor=blue]{hyperref}

\begin{document}
%TC:ignore

%\title{State-selective laser ionization facilitates direct mass measurements of the high-spin isomers in self-conjugate $^{94}$Ag}
\title{Direct mass measurements of the high-spin isomers in self-conjugate $^{94}$Ag}

\author{{Ville} {Virtanen}}\email{ville.a.virtanen@jyu.fi}\thanks{These authors contributed equally to this work.}
\affiliation{Department of Physics, Accelerator laboratory, University of Jyväskylä, P.O. Box 35 (YFL), Jyväskylä, FI-40014, Finland}
\author{{Mikael} {Reponen}}\email{mikael.h.t.reponen@jyu.fi}\thanks{These authors contributed equally to this work.}
\affiliation{Department of Physics, Accelerator laboratory, University of Jyväskylä, P.O. Box 35 (YFL), Jyväskylä, FI-40014, Finland}

\author{{Duy Duc} {Dao}}
\affiliation{{Universit\'e de Strasbourg}, {CNRS}, {IPHC UMR 7178}, {{23 rue du Loess, Strasbourg }, {F-67000}, {France}}}
\author{{Frédéric} {Nowacki}}
\affiliation{{Universit\'e de Strasbourg}, {CNRS}, {IPHC UMR 7178}, {{23 rue du Loess, Strasbourg }, {F-67000}, {France}}}

\author{{Tommi} {Eronen}}
\affiliation{Department of Physics, Accelerator laboratory, University of Jyväskylä, P.O. Box 35 (YFL), Jyväskylä, FI-40014, Finland}

\author{{Anu} {Kankainen}}
\affiliation{Department of Physics, Accelerator laboratory, University of Jyväskylä, P.O. Box 35 (YFL), Jyväskylä, FI-40014, Finland}

% Please check if anyone is missing
\author{{Olga} {Beliuskina}}
\affiliation{Department of Physics, Accelerator laboratory, University of Jyväskylä, P.O. Box 35 (YFL), Jyväskylä, FI-40014, Finland}

\author{{Lucia} {Caceres}}
\affiliation{{Grand Accélérateur National d’Ions Lourds (GANIL)}, {CEA/DRF-CNRS/IN2P3}, {{Bvd Henri Becquerel}, {Caen}, {14076}, {France}}}
\author{{Saikumar} {Chinthakayala}}
\affiliation{Department of Physics, Accelerator laboratory, University of Jyvaskyla, P.O. Box 35 (YFL), Jyväskylä, FI-40014, Finland}
\affiliation{{Grand Accélérateur National d’Ions Lourds (GANIL)}, {CEA/DRF-CNRS/IN2P3}, {{Bvd Henri Becquerel}, {Caen}, {14076}, {France}}}

\author{{Ruben P.} {de Groote}}
\affiliation{{Instituut voor Kern- en Stralingsfysica}, {KU Leuven}, {{Leuven}, {B-3001}, {Belgium}}}

\author{{Zhuang} {Ge}}
\affiliation{Department of Physics, Accelerator laboratory, University of Jyväskylä, P.O. Box 35 (YFL), Jyväskylä, FI-40014, Finland}
\affiliation{{GSI Helmholtzzentrum f{\"u}r Schwerionenforschung GmbH}, {{Darmstadt}, {64291}, {Germany}}}

\author{{Jason D.} {Holt}}
\affiliation{{TRIUMF}, {{Vancouver}, { 4004 Wesbrook Mall, BC V6T 2A3}, {Canada}}}
\affiliation{{Department of Physics}, {McGill University}, {{Montr\'eal}, {QC H3A 2T8}, {Canada}}}

\author{{Baishan} {Hu}}
\affiliation{{Cyclotron Institute and Department of Physics and Astronomy}, {Texas A\&M University}, {{College Station}, {Texas 77843}, {USA}}}

\author{{Arthur} {Jaries}}
\affiliation{Department of Physics, Accelerator laboratory, University of Jyväskylä, P.O. Box 35 (YFL), Jyväskylä, FI-40014, Finland}
\affiliation{{ European Organization for Nuclear Research (CERN)}, {{1211 Geneva 23 }, {Switzerland}}}

\author{{Brian} {Kootte}}
\affiliation{Department of Physics, Accelerator laboratory, University of Jyvaskyla, P.O. Box 35 (YFL), Jyväskylä, FI-40014, Finland}

\author{{Gabriella} {Kripk{\'o}-Koncz}}
\affiliation{{II Physikalisches Institut}, {Justus-Liebig-Universit\"at Gie\ss en}, { {Gie\ss en}, {35392}, {Germany}}}
\affiliation{{School of Physics and Astronomy}, {University of Edinburgh}, {  {EH9 3FD Edinburgh}, {United Kingdom}}}

\author{{Sonja} {Kujanpää}}
\affiliation{Department of Physics, Accelerator laboratory, University of Jyväskylä, P.O. Box 35 (YFL), Jyväskylä, FI-40014, Finland}
\affiliation{{Laboratory for Biological Geochemistry, School of Architecture, Civil \& Environmental Engineering, École Polytechnique Fédérale de Lausanne}, {  {GR C2 505, Station 2, 1015 Lausanne}, {Switzerland}}}

\author{{Gwendal} {Loaëc}}
\affiliation{{ENSICAEN}, {  {6 Bd Maréchal Juin Caen}, {France}}}

\author{{Iain D.} {Moore}}
\affiliation{Department of Physics, Accelerator laboratory, University of Jyväskylä, P.O. Box 35 (YFL), Jyväskylä, FI-40014, Finland}

\author{{Luis Miguel} {Motilla}}
\affiliation{Department of Physics, Accelerator laboratory, University of Jyväskylä, P.O. Box 35 (YFL), Jyväskylä, FI-40014, Finland}
\affiliation{{Universit\'e de Strasbourg}, {CNRS}, {IPHC UMR 7178}, {{23 rue du Loess, Strasbourg }, {F-67000}, {France}}}

\author{{Maxime} {Mougeot}}
\affiliation{Department of Physics, Accelerator laboratory, University of Jyväskylä, P.O. Box 35 (YFL), Jyväskylä, FI-40014, Finland}

\author{{Stylianos} {Nikas}}
\affiliation{Department of Physics, Accelerator laboratory, University of Jyväskylä, P.O. Box 35 (YFL), Jyväskylä, FI-40014, Finland}

\author{{Ilkka} {Pohjalainen}}
\affiliation{Department of Physics, Accelerator laboratory, University of Jyväskylä, P.O. Box 35 (YFL), Jyväskylä, FI-40014, Finland}

\author{{Andrea} {Raggio}}
\affiliation{Department of Physics, Accelerator laboratory, University of Jyväskylä, P.O. Box 35 (YFL), Jyväskylä, FI-40014, Finland}
\affiliation{{Instituut voor Kern- en Stralingsfysica}, {KU Leuven}, {{Leuven}, {B-3001}, {Belgium}}}

\author{{Wirunchana} {Rattanasakuldilok}}
\affiliation{Department of Physics, Accelerator laboratory, University of Jyväskylä, P.O. Box 35 (YFL), Jyväskylä, FI-40014, Finland}

\author{{Nicolas} {Richterich}}
\affiliation{Department of Physics, Accelerator laboratory, University of Jyväskylä, P.O. Box 35 (YFL), Jyväskylä, FI-40014, Finland}
\affiliation{{School of Electrical Engineering, Aalto University}, {  {Maarintie 8, Espoo, 00076}, {Finland}}}

\author{{Jouni} {Ruotsalainen}}
\affiliation{Department of Physics, Accelerator laboratory, University of Jyväskylä, P.O. Box 35 (YFL), Jyväskylä, FI-40014, Finland}

\author{{Marek} {Stryjczyk}}
\affiliation{Department of Physics, Accelerator laboratory, University of Jyväskylä, P.O. Box 35 (YFL), Jyväskylä, FI-40014, Finland}
\affiliation{{Institut Laue-Langevin}, {  {71 Avenue des Martyrs, F-38042 Grenoble}, {France}}}

\author{{Juha} {Äystö}}
\affiliation{Department of Physics, Accelerator laboratory, University of Jyväskylä, P.O. Box 35 (YFL), Jyväskylä, FI-40014, Finland}

\date{October 2026}

\begin{abstract}
Two decades ago, two-proton emission from a (21$^+$) isomeric state in $^{94}$Ag was reported. This report has remained the sole example of possible two-proton radioactivity in a nucleus with an odd number of protons on the chart of nuclei. A debate regarding the viability of the decay mode has been ongoing due to the unknown mass of the isomer. Here, using state-selective laser ionization in a hot cavity catcher, we present multi-reflection time-of-flight mass spectrometry of the long-lived isomeric states of $^{94}$Ag and its one-proton decay daughter $^{93}$Pd. For the first time, direct mass measurements of the (7$^+$) and (21$^+$) isomeric states set the stage for an independent discussion of the energetics of these decays. We rule out the two-proton emission from the (21$^+$) state as previously reported and reveal the one-proton decay energy to be significantly lower. State-of-the-art shell-model calculations reproduce the energy gap of the (7$^+$) and the (21$^+$) states as well as the half-life of the (21$^+$) state, implying the (21$^+$) state is likely near-spherical, contradicting the reported explanation for the two-proton radioactivity. Importantly, the techniques employed are a stepping stone towards the first laser spectroscopic investigation of a proton-emitting state.
\end{abstract}
%TC:endignore
\maketitle
% Two-proton radioactivity~\cite{delion_two-proton_2022, pfutznerTwoprotonEmissionRelated2023} was first considered by Zel'dovich~\cite{zeldovichEXISTENCENEWISOTOPES1960} and Goldansky in 1960~\cite{Goldansky1960}. This type of decay is generally described as a sequential two-body process or a three-body process \cite{Blank_2008}. These decays occur very rarely in contrast to the relatively common one-proton radioactivity~\cite{kokkonen_new_2025, zhang_observation_2022}. The radioactive decay by two-proton-emission has so far been reliably observed either in a few extremely short-lived species~\cite{pfutznerTwoprotonEmissionRelated2023} or in heavier nuclei near the proton dripline, for example $^{45}\textrm{Fe}$~\cite{Pfutzner2002,Giovinazzo2002}, $^{48}\textrm{Ni}$~\cite{PhysRevC.83.061303}, $^{54}\textrm{Zn}$~\cite{Blank2005} and $^{67}\textrm{Kr}$~\cite{goigoux_two-proton_2016}. Notably, each of these nuclei has an even number of protons. In 2006, unexpected two-proton radioactivity was reported from a high-spin (21$^+$) isomer of the $N = Z = 47$ nucleus $^{94}$Ag~\cite{Mukha2006}. The isomer is a unique $\beta$-decaying state that is highly unbound against one- and two-proton decay~\cite{Mukha2004}. The reported decay channel~\cite{Mukha2006} would make it so far the only odd proton-number nuclear state to undergo two-proton decay \cite{Aysto2006}. The possibility of this decay was questioned, leading to an intense debate~\cite{Pechenaya2007, Mukha2008a,Pechenaya2008} regarding its existence~\cite{Jenkins_reviewing_2009}.
Two-proton radioactivity~\cite{delion_two-proton_2022, pfutznerTwoprotonEmissionRelated2023} was first considered by Zel'dovich~\cite{zeldovichEXISTENCENEWISOTOPES1960} and Goldansky in 1960~\cite{Goldansky1960}. This type of decay is generally described as either a sequential two-body process or a three-body process \cite{Blank_2008}, occurring very rarely in contrast to the relatively common one-proton radioactivity~\cite{kokkonen_new_2025, zhang_observation_2022}. Radioactive decay by two-proton emission has so far been reliably observed in a few extremely short-lived species as well as in heavier nuclei near the proton dripline~\cite{pfutznerTwoprotonEmissionRelated2023}. Notably, each of these nuclei has an even number of protons. In 2006, two-proton radioactivity was unexpectedly reported from a high-spin (21$^+$) isomer of the $N = Z = 47$ nucleus $^{94}$Ag~\cite{Mukha2006}. The nucleus has a ($0^+$) ground state, and two long-lived isomers, (7$^+$) and (21$^+$). The (21$^+$) isomer is a unique $\beta$-decaying state that is highly unbound against one- and two-proton decay~\cite{Mukha2004}. The reported decay channel~\cite{Mukha2006} would make it the only odd proton-number nuclear state to undergo two-proton decay \cite{Aysto2006}. The possibility of this decay has, however, been questioned by subsequent experiments~\cite{Pechenaya2007, Kankainen2008, Cerny2009, gabi93pd_accepted} and theoretical efforts~\cite{Kaneko2008, aggarwalProtonRadioactivityNoncollective2010}, leading to a continued debate~\cite{Pechenaya2007, Mukha2008a,Pechenaya2008} regarding its existence~\cite{Jenkins_reviewing_2009}.
% The reported $_{47}^{94}\mathrm{Ag} (21^+) \rightarrow_{45}^{92}\mathrm{Rh}^* + 2\mathrm{p}$ decay signature was a 1.9 MeV sum-energy peak of two charged particles detected with separate silicon detectors in coincidence with gamma-ray transitions known to de-excite the high-spin states in $^{92}$Rh~\cite{Mukha2006}. However, an experiment to confirm the two-proton radioactivity saw no evidence of the 1.9~MeV sum-energy peak and could only verify the 0.79~MeV one-proton decay branch~\cite{Cerny2009}. Furthermore, high-precision nuclear mass data contradict the original decay data~\cite{Kankainen2008, gabi93pd_accepted}, and some of the eight $\gamma$-ray transitions attributed to $^{92}$Rh in~\cite{Mukha2006} were not found in a subsequent in-beam gamma-ray spectroscopy study~\cite{Pechenaya2007}, indicating the two-proton emission may lead to states below the  (11$^+$) state in $^{92}$Rh. 
% Moreover, the two-proton emission was explained to contend with other decay modes such as $\beta$-decay, $\beta$-delayed proton emission~\cite{Plettner2004,Mukha2004} and one-proton radioactivity~\cite{Mukha2005} due to a large prolate deformation of the $(21^+)$ state. Theoretical calculations~\cite{Kaneko2008, aggarwalProtonRadioactivityNoncollective2010} have not supported this explanation, however. \textbf{Laser ionization schemes and time-of-flight spectra.} 
\begin{figure*}[hbt!]
  \centering
  \includegraphics[width=\textwidth]{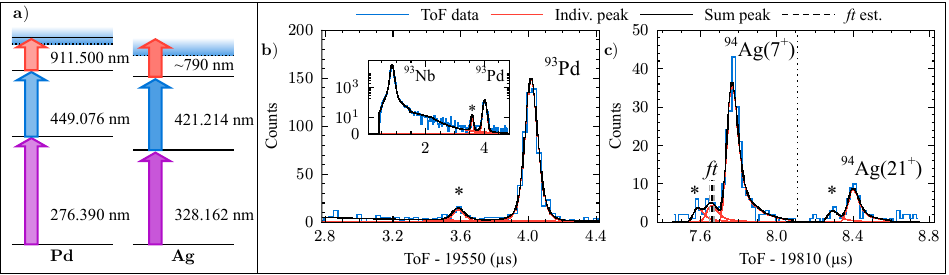}
  \caption{Isotopes of Pd and Ag were selected by resonantly laser ionizing them, as shown in panel  \textbf{(a)}. The element-selective nature of this process was important for the identification of $^{93}$Pd and the states in $^{94}$Ag measured with the MR-ToF-MS. These spectra are shown in  \textbf{(b)} and \textbf{(c)}. The time-of-flight spectrum of $^{94}$Ag in \textbf{(c)} consists of two different measurements, marked by the dotted line at 8.1 $\mu$s, where the laser frequencies are tuned to either ionize the (7$^{+}$)  or the (21$^+$) state. The vertical dashed lines mark the expected time-of-flight for the $^{94}$Ag ground state based on the superallowed $\beta$-decay $ft$ value. The $*$  marks contaminant peaks which show no statistically significant laser response. \label{fig: setup}}
\end{figure*}

Proton radioactivity may occur when the daughter nucleus is more bound than the decaying parent state. This binding energy difference, known as the decay $Q$ value, can be determined from the masses of the initial and final-state nuclei, and it must be large enough for the proton emission to compete with other decay modes, such as $\beta$-decay~\cite{pfutznerRadioactiveDecaysLimits2012}. To explore if the two-proton decay mode is energetically feasible, it is thus crucial to directly measure the masses of $^{92}$Rh and $^{94}$Ag (21$^+$). Although the mass of the two-proton decay daughter $^{92}$Rh is  known precisely~\cite{weberMassMeasurementsVicinity2008, fallisMassMeasurementsIsotopes2011, Kankainen2008, gabi93pd_accepted}, the masses of the long-lived states in $^{94}$Ag have remained unmeasured. So far, knowledge on the mass excess of the (21$^+$) state has been based on the one-proton decay $Q$ value and the mass of $^{93}$Pd~\cite{NUBASE20,Mukha2005,Cerny2009}, whereas the mass of $^{93}$Pd has, until a recent direct measurement~\cite{gabi93pd_accepted}, been based on an experimental $\beta$ decay $Q$ value~\cite{parkNewComprehensiveEnsuremath2019, wangAME2020Atomic2021}. With the recent direct mass measurement of $^{93}$Pd, the mass of $^{94}$Ag $(21^+)$, deduced from the single proton energies, remains in a 1.090(110)~MeV contradiction with the $-$44.960(100)~MeV mass excess determined from the 1.9(1)~MeV two-proton decay energy~\cite{gabi93pd_accepted}. Therefore, a direct mass measurement of the $^{94}$Ag (21$^+$) state is vital to resolve the debate on the two-proton decay from the odd-odd \textit{N = Z} nucleus~$^{94}$Ag.

In this Letter, we report the first mass measurements of the isomeric (7$^+$) and (21$^+$) states of $^{94}$Ag, in addition to measuring the mass of $^{93}$Pd, using a novel hot cavity catcher laser ion source and a Multi-Reflection Time-of-Flight Mass Spectrometer (MR-ToF-MS)~\cite{virtanenHighresolutionMultireflectionTimeofflight2025}. We also estimate the mass of the (0$^+$) ground state from $\beta$-decay $ft$ values.  We have obtained with the measured masses the one- ($Q_p$) and two-proton ($Q_{2p}$) decay $Q$ values for the (21$^+$) isomer and pin down the one- and two-proton decay channels, using state-of-the-art nuclear theory calculations to further elucidate the properties of the unique (21$^+$) isomer.
% Our work brings to conclusion the long-standing debate on the claimed two-proton radioactivity of the $^{94}$Ag~(21$^+$) isomer. 

These measurements were performed at the Ion Guide Isotope Separation On-Line (IGISOL) facility~\cite{moore_igisol_2014} of the University of Jyväskylä Accelerator Laboratory. Both $^{94}$Ag and $^{93}$Pd were produced by impinging an approximately 5~MeV per nucleon beam of $^{40}$Ca on a thin $^{58}$Ni target foil. The products were element-selectively laser ionized  within the hot cavity catcher laser ion source, extracted and accelerated to 30~keV, mass separated, and injected into the radiofrequency quadrupole cooler-buncher \cite{NIEMINEN2001244,virtanenMiniaturisedCoolerbuncherReduction2025}. Finally, the ion masses were determined based on their time-of-flights in the MR-ToF-MS using the $C_\mathrm{ToF}$ method \cite{Wienholtz2013}. Additional identification of $^{94}$Ag and $^{93}$Pd was made via the requirement for all resonant steps to be present in the resonant laser ionization process. The three-step excitation and ionization schemes are indicated in Figure \ref{fig: setup}. The large hyperfine splitting in the atomic states of silver provided an additional means to distinguish between the (7$^+$) and (21$^+$) nuclear states via partial isomer-selective ionization. The use of lasers allowed us to differentiate the mass peaks of the isotopes of interest from background peaks. While not all of the background features were identified, they were included in the fitting procedure when they were close to the masses of interest. The setup facilitated measurements with rates as low as three detected $^{94}$Ag~(21$^+$) ions per hour. Further details are presented in the End Matter.  

% of the experimental setup and data analysis procedures

%-----------------------------Table 1 --------------------------------
%\textbf{Experimental mass excess data.}
\begin{table*}[hbt!]

% \caption[Mass excess for $^{94}$Ag$(7^+, 21^+)$, $^{93}Pd$, $^{93}Nb$, $^{94}$Zr and $^{95}$Ag]{\textbf{Experimental mass excess data} Mass excess for $^{94}$Ag$(7^+, 21^+)$, $^{93}$Pd, $^{93}$Nb, $^{94}$Zr and $^{95}$Ag from this work. The literature values are taken from AME2020~\cite{wangAME2020Atomic2021} with the exception of $^{95}$Ag, which is from the recent JYFLTRAP measurement~\cite{ge2024highprecision}, marked with $*$, and $^{93}$Pd, which is from an experiment performed at the FRS Ion Catcher, GSI~\cite{gabi93pd_accepted}, marked with $\dagger$. We note that the measurement of $^{94}$Zr was performed in a separate run and was used as a reliability check of the experimental method. A higher statistics measurement of $^{94}$Zr was used as a reference for $^{93}$Pd and $^{93}$Nb. The residual measurement uncertainty was estimated from the difference to well-known stable nuclei and found to be approximately 30 keV (in square brackets). It has been treated as a common systematic uncertainty. The \# marks an AME extrapolated value. \label{tab: massexesses}}
\caption[Mass excess of $^{94}$Ag$(7^+, 21^+)$, $^{93}Pd$, $^{93}Nb$, $^{94}$Zr and $^{95}$Ag]{ 
The mass excesses (ME) of $^{94}$Ag$(7^+, 21^+)$, $^{93}$Pd, $^{93}$Nb, $^{94}$Zr and $^{95}$Ag nuclei, with their spin-parities given in the $J^\pi$ column, were determined via $C_\mathrm{ToF}$ measurements which utilized pairs of reference nuclei given in the Refs. -column. The results are compared with  AME2020~\cite{wangAME2020Atomic2021} values or data from recent precision measurements on $^{93}$Pd (a) and $^{95}$Ag (b). We note that the measurement of $^{94}$Zr was performed in a separate run and was used as a reliability check of the experimental method. A higher statistics measurement of $^{94}$Zr was used as a reference for $^{93}$Pd and $^{93}$Nb. The residual measurement uncertainty was estimated from the difference to well-known stable nuclei and found to be approximately 30 keV (in square brackets). It has been treated as a common systematic uncertainty. The \# marks an AME extrapolated value. \label{tab: massexesses}}
\centering

\begin{tabular}[b]{lcccclr}
 %-59 127(35)~keV
\hline
\hline
Nuclide & $J^\pi$ & Refs.& $C_\mathrm{ToF}$ & ME (keV)  & ME$_\mathrm{Lit.}$ (keV) & ME-ME$_\mathrm{Lit.}$ (keV) \\

\hline

%$^{94}$Ag & $(0^+)$&$^{94}$Mo, $^{96}$Mo& xx.XXXX (XX)& -53 720 (160)&  -52 400 (400)\# & -1320\# (430\#) \\
$^{94}$Ag& $(7^+)$&$^{94}$Mo, $^{96}$Mo& 0.480616(15)& $-$52 496(28)[30]&  $-$51 050(570)\# & $-$1400(600)\# \\
$^{94}$Ag& $(21^+)$ & $^{94}$Mo, $^{96}$Mo& 0.477660(50)& $-$47 010(90)[30]&  $-$45 900(370) &$-$1100(400) \\
% \hline
% $^{93}$Pd& $9/2^+$ & $^{92}$Zr, $^{94}$Zr & 
% $-$0.016734(7)& $-$59 177(13)[30] &  $-$58 980(370) & $-$200(380) \\
\hline
$^{93}$Pd& $9/2^+$ & $^{92}$Zr, $^{94}$Zr & 
$-$0.016734(7)& $-$59 177(13)[30] &   $-$59 127(35)\footnote{From an experiment performed at the FRS Ion Catcher, GSI~\cite{gabi93pd_accepted}} & $-$50(48) \\
\hline
$^{93}$Nb& $9/2^+$ & $^{92}$Zr, $^{94}$Zr & $-$0.0017074(11) & $-$87 191(2)[30] & $-$87 212.8(15)&  22(30)\\
$^{95}$Ag& $9/2§^+$ & $^{95}$Mo, $^{96}$Mo& 0.469840(11)& $-$59 722(10)[30] &  $-$59 743.3(14)\footnote{From an experiment performed at the JYFLTRAP~\cite{ge2024highprecision}}  & 21(32) \\
$^{94}$Zr& $0^+$ & $^{94}$Mo, $^{96}$Mo& 0.499360(50)& $-$87 230(90)[30]&  $-$87 269.33(16)  &40(100) \\

%$^{93}$Pd& $\frac{9}{2}^+$&  $^{92}$Zr, $^{94}$Zr& ? & ? & ?  \\
\hline
\hline
\end{tabular}
\end{table*}
The $^{93}$Pd, $^{94}$Ag (7$^+$) and (21$^+$) peaks, clearly shown in the time-of-flight spectra of Fig.~\ref{fig: setup} (b), were the only statistically significant features to respond to laser parameters, allowing us to  reliably identify these ion species and to determine their masses. The results are summarized in Table~\ref{tab: massexesses} where we also demonstrate the reliability of the experimental method by comparing measured masses of $^{93}$Nb, $^{94}$Zr and $^{95}$Ag to existing high-precision data~\cite{ge2024highprecision,gulyuzDeterminationDirectDoublensuremathbetadecay2015}. Our results match the literature values within $\approx 3\times10^{-7}$ on average, i.e., 30~keV -- typical for MR-ToF mass measurements \cite{Wienholtz2013,Ito2018,Reiter2018,Ayet2019}. This empirically deduced 30~keV accuracy is incorporated to the error budget. A more detailed breakdown will be published elsewhere.  

Our mass excess results for $^{94}$Ag show that the (7$^+$) and (21$^+$) states are 1.4~MeV and 1.1~MeV more bound, respectively, than literature suggests~\cite{NUBASE20} and that the measured mass of the (21$^+$) state significantly differs from the recent estimates~\cite{gabi93pd_accepted}. The one-proton decay daughter $^{93}$Pd is also found to be 0.2~MeV more bound compared to AME2020~\cite{wangAME2020Atomic2021} -- but still within the  uncertainty of the previous experimental value \cite{parkNewComprehensiveEnsuremath2019} -- and in close agreement with the recent direct measurement from GSI~\cite{gabi93pd_accepted}. These results furthermore suggest that $^{94}$Ag (0$^+$) should be approximately 1~MeV more bound than the previously estimated value, $-52400 (400)$\#~keV~\cite{wangAME2020Atomic2021}. 

To estimate the mass of the (0$^+$) ground state of $^{94}$Ag, we utilized its superallowed Fermi $\beta$-decay to the ground state of $^{94}$Pd. Using a weighted mean of the measured half-life values: 27(2)~ms \cite{parkNewComprehensiveEnsuremath2019} and 26(1)~ms \cite{Appleton2025}, an estimated 98.68~\% branching ratio \cite{Mare2022}, and the superallowed $\beta$-decay $ft$ value of the nearby $^{74}$Rb, 3082.8(65)~s \cite{Hardy2020}, we obtain $Q_{EC}=12.670(100)$~MeV for the $\beta$-decay, assuming the same transition-dependent radiative and nuclear-structure corrections as for $^{74}$Rb ~\cite{Hardy2020}. Combined with the $^{94}$Pd mass~\cite{wangAME2020Atomic2021}, the $Q_{EC}$ value yields a ground-state mass excess of $-53430(100)$~keV for $^{94}$Ag in agreement with Ref.~\cite{gabi93pd_accepted}. We have labeled the expected time-of-flight range of $^{94}$Ag(0$^+$) to Fig.~\ref{fig: setup} (b) with $ft$ -marker. We note that there are counts in the time-of-flight region that match with the predicted ground-state mass (see Fig.~\ref{fig: setup}); however, the statistics are insufficient to declare the ground state identified.
% \begin{figure*}
%     \centering
%     \includegraphics[width=1\textwidth]{Figures/levelscheme.pdf}
%     \caption[]{\textbf{One- and two-proton decay schemes for $^{94}$Ag(21$^+$).} \textbf{(a)} The (21$^+$) isomer in $^{94}$Ag has a positive two-proton decay $Q$ value, $Q_{\mathrm{2p}}=1.41(10)$~MeV, and therefore can only populate states with spins less or equal to $10~\hbar$ in the daughter nucleus $^{92}$Rh. The $Q_{2p}$ determined in \cite{Mukha2006}, 1.9(1)~MeV, is energetically impossible from the (21$^+$) state. Based on the new mass excess value of $^{93}$Pd, the one-proton decay branches (red arrows) from Ref.~\cite{Mukha2005} would lead to states between 3.7-4.1 MeV in the daughter nucleus $^{93}$Pd marked in blue. The orange boxes indicate the uncertainty in the populated levels, including the uncertainty of the $^{94}$Ag(21$^+$) mass excess and for the one-proton emission, also the uncertainties of the proton energies.  Here, the isomer excitation energies in $^{94}$Ag are based on the estimated $(0^+)$ ground-state mass. Only relevant states are presented for $^{92}$Rh and $^{93}$Pd; \textbf{I} and \textbf{II} indicate the positive and negative parity bands in $^{93}$Pd, respectively. Energy values are given in MeV. \textbf{(b)} The measured $^{94}$Ag(21$^+$) mass excess compared to the mass excesses calculated from the single-proton data combined with the energetically allowed states of $^{93}$Pd (blue markers). \label{fig:decayscheme}}
% \end{figure*} \textbf{One- and two-proton decay schemes for $^{94}$Ag(21$^+$).} 
\begin{figure*}
    \centering
    \includegraphics[width=1\textwidth]{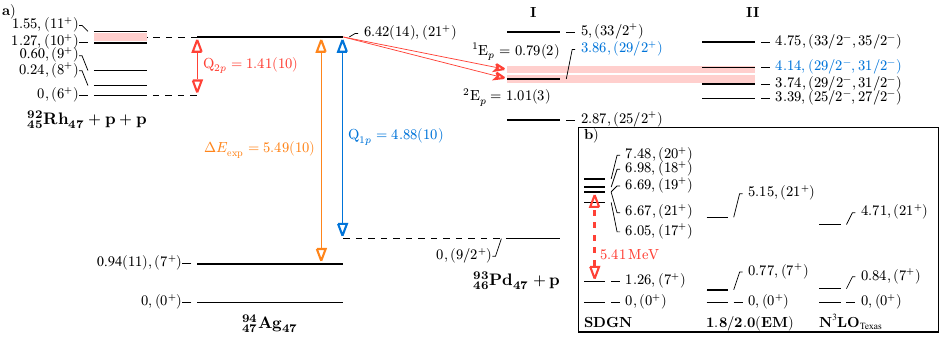}
    \caption[]{\textbf{(a)} The (21$^+$) isomer in $^{94}$Ag has a positive two-proton decay $Q$ value, $Q_{\mathrm{2p}}=1.41(10)$~MeV, and therefore can only populate states with spins less or equal to $10~\hbar$ in the daughter nucleus $^{92}$Rh. The $Q_{2p}$ determined in \cite{Mukha2006}, 1.9(1)~MeV, is energetically impossible. Based on the new mass excess value of $^{93}$Pd, the one-proton decay branches (red arrows) from Ref.~\cite{Mukha2005} would lead to states in the daughter nucleus $^{93}$Pd marked in blue. The pink boxes indicate the uncertainty in the populated levels and it includes the uncertainty of the $^{94}$Ag(21$^+$) isomer mass excess and the uncertainties of the proton energies. Here, the isomer excitation energies in $^{94}$Ag are based on the estimated $(0^+)$ ground-state mass. Only relevant states are presented for $^{92}$Rh and $^{93}$Pd; \textbf{I} and \textbf{II} indicate the positive and negative parity bands in $^{93}$Pd, respectively. Energy values are given in MeV. \textbf{(b)} The experimental energy difference between the (7$^+$) and (21$^+$) isomers in $^{94}$Ag, marked with orange arrow in \textbf{(a)}, is reproduced only by the LSSM calculations utilizing the SDGN effective interaction (Dashed red arrow) within experimental uncertainties.  \label{fig:decayscheme}}
\end{figure*}
%$^{94}$Ag(21$^+$) mass excess and for the one-proton emission, also the uncertainties of the proton energies. 
%between 3.7-4.1 MeV 
% . VS-IMSRG calculations with 1.8/2.0 (EM), and N$^3$LO$_{\rm Texas}$ chiral interactions produce a significantly lower excitation energy for the (21$^+$) state, and thus a smaller energy gap between it and the (7$^+$) state. In addition, only the LSSM calculations reproduce the mechanism for the (21+) spin-gap isomerism
%experimental energy difference between the (7$^+$) and (21$^+$) isomers in 94Ag .
%Here the first states occurring are presented.
We have determined the single-proton and two-proton decay $Q$ values using our new mass values for $^{93}$Pd and $^{94}$Ag (21$^+$), and $^{92}$Rh from AME2020~\cite{wangAME2020Atomic2021}. The one-proton decay energy $Q_{1p}=4.878(100)$~MeV is substantially lower than the previously reported $Q_{1p}=5.78(3)$~MeV \cite{Mukha2005} value. This constrains the $^{93}$Pd states to which the observed $0.79$~MeV and \mbox{1.01~MeV} proton branches \cite{Mukha2005} may proceed. In the relevant energy ranges, determined by the measured proton energies and the mass of the $(21^+)$ state (see Fig.~\ref{fig:decayscheme}), two candidate states in $^{93}$Pd are identified: a $(29/2^+)$  state at a $3.8632$~MeV excitation energy and a \mbox{($29/2^-$, $31/2^-$)} state at $4.1387$~MeV. The $(29/2^+)$ state could connect to $^{94}$Ag$(21^+)$ by an $l_\mathrm{p} = 6$~$\hbar$ proton and, likewise, the \mbox{($29/2^-$, $31/2^-$)} state could be connected by an \mbox{$l_\mathrm{p} = 5$~$\hbar$ or $7$~$\hbar$} proton, analogous to the 1p decay of $^{53}$Co$(19/2^-)$ \cite{sarmientoElucidatingNatureProton2023}.

Combining the excitation energies of these candidate states with the mass of $^{93}$Pd and the single-proton energies gives a $-46 960(40)$~keV mass excess for the initial state of the 0.79(2)~MeV proton branch, whereas the second, 1.01(3)~MeV proton branch corresponds to an initial state with a  mass excess of $-47 010(50)$~keV. These values are consistent with each other and with the measured mass excess of the (21$^+$) state in $^{94}$Ag. However, the level scheme of $^{93}$Pd, including the spins and parities of the candidate states, is somewhat uncertain \cite{Baglin2011}; therefore, the $\gamma$ rays associated with the proton decay should be investigated using high-purity ion beams of $^{94}$Ag to reconcile the candidate states with some of the $\gamma$-ray transitions proposed to coincide with the proton branches~\cite{Mukha2005}.

With the newly determined mass value for the (21$^+$) isomer of $^{94}$Ag, we obtain Q$_{\mathrm{2p}}=1.41(10)$~MeV (see Fig.~\ref{fig:decayscheme}). It is therefore energetically impossible that the 1.9(1)~MeV sum-energy peak observed in Ref.~\cite{Mukha2006} would have been produced by two-proton emission from the (21$^+$) isomer of $^{94}$Ag. Although the Q$_{\mathrm{2p}}$ value remains positive, due to its value, the two-proton decay may feed only the (6$^+$), (8$^+$), (9$^+$) or (10$^+$) states in $^{92}$Rh and would involve a spin difference of $ L=11~\hbar-15~\hbar$ with respect to the initial (21$^+$) state, in agreement with the conclusions of Ref.~\cite{Pechenaya2007} and in contrast to conclusions in~\cite{Mukha2006}, where $L=6~\hbar-10~\hbar$ was assumed. 

While measurements of the quadrupole moment are needed to probe the shape of the (21$^+$) isomer, we can test the reliability of nuclear theory predictions with our new mass data. State-of-the-art shell-model and ab-initio valence-space in-medium similarity renormalization group (VS-IMSRG)~\cite{Stroberg2019} methods have been recently applied to calculate the excitation energies of nearby low-lying, low-spin isomers of silver, indium and rhodium isotopes around the $N$=50 shell closure~\cite{hornung_isomer_2020, ge2024highprecision, niesIsomericExcitationEnergy2023, mollaebrahimiStudyingGamowTellerTransitions2023}. Here, we extend the methods to elucidate the properties of the (21$^+$) isomer. We have evaluated the binding energy of $^{94}$Ag and the $\beta$-decay half-life of its (21$^+$) state using the shell-model approach and estimated the shape of the (21$^+$) isomer with the Discrete Non-Orthogonal Shell-Model (DNO-SM) formalism~\cite{daoNuclearStructureDiscrete2022,daoNuclearStructureDiscrete2025}. We have performed large-scale shell-model (LSSM) calculations using the SDGN effective interaction~\cite{nowackiShellModelDescription2002}, and VS-IMSRG calculations using 1.8/2.0 (EM)~\cite{Simon2017}  and N$^3$LO$_{\rm Texas}$~\cite{hu2025neutrondriplinecalciumisotopes} chiral interactions (see End Matter). The shell-model calculations reproduce the spin-gap isomerism of the (21$^+$)  state and estimate the studied (7$^+$) and (21$^+$) isomeric states to differ by 5.41~MeV in energy, excellently agreeing with the  5.49(10)~MeV experimental value (see Fig.~\ref{fig:decayscheme} (b)). The ab-initio calculations do not reproduce the spin-gap isomerism nor estimate the energy difference with the same accuracy as the LSSM method, giving 4.38~MeV with 1.8/2.0 (EM)  and 3.87~MeV with N$^3$LO$_{\rm Texas}$. We estimate an uncertainty of 1~MeV arising from the decoupling of different valence spaces, which underlines the need for further improvements~\cite{He2024}.
%\textbf{Theoretical calculations.} 
\begin{figure*}
\centering
\includegraphics[width=1\linewidth]{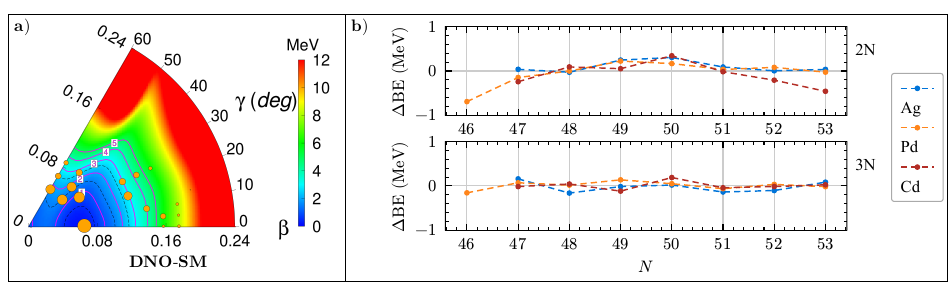}
\caption{\textbf{(a)} The structure of $^{94}$Ag$(21^+)$ represented by a ($\beta$, $\gamma$) potential energy surface calculated with the DNO-SM formalism. The orange circles map the contribution of constrained Hartree-Fock basis vectors. \textbf{(b)} Difference between experiment and shell model based binding energies $\Delta$BE using two-nucleon (2N) and three-nucleon (3N) master terms for Pd, Ag and Cd. The experimental mass data is either from AME2020 except for the data in Table \ref{tab: massexesses} and the estimated mass of $^{94}$Ag. }\label{fig:sm94Ag}
\end{figure*}
% This emphasizes the suitability of the LSSM method in describing nuclear properties in the region.
The potential energy surface presented in  Fig.~\ref{fig:sm94Ag}(a) illustrates the sphericity of the (21$^+$) isomer with wave function components developed in the spherical region $\beta \sim 0-0.1$ and with no evidence for an intrinsic deformed structure. This feature agrees with other theoretical estimates from Refs.~\cite{Kaneko2008, aggarwalProtonRadioactivityNoncollective2010} and contradicts the large deformation claimed in Ref. \cite{Mukha2006}. In the present case, the initial (21$^+$) state and the final yrast (20$^+$) state in $^{94}$Pd arise from the spin cut-off of the single (g$_{\frac{9}{2}}$)$^{14}$ configuration. Due to the large \textit{N=Z=}50 shell gap, this configuration represents the dominant component in the full valence space, up to 72\% and 74\% in $^{94}$Ag and  $^{94}$Pd, respectively. As a consequence, the decay of the $(21^+)$ state  mainly proceeds via a single $\beta$~-decay branch between these states, and the half-life is directly sensitive to the $Q$ value. Using the newly measured mass excess of the $(21^+)$ state, we find a $\beta$-decay half-life of 0.45(3)~s following the procedure described in the End Matter. This matches with the measured half-life, 0.40(4)~s~\cite{NDS94}, whereas the 0.23(3)~s half-life calculated based on the literature mass~\cite{NUBASE20} is clearly too short. 

Figure~\ref{fig:sm94Ag}(b) presents the shell-model evaluated binding energies of the neighboring palladium, silver and cadmium chains between 46 $\leq$ N $\leq$ 53. The results closely agree with the combined AME2020 and the data from this work (Table \ref{tab: massexesses}), deviating only $0.251$~MeV and $0.115$~MeV rms for quadratic and cubic master terms, respectively. Notably, if the $-52400 (400)$\#~keV AME2020 value for $^{94}$Ag mass is used instead of the $-53430(100)$~keV estimate from this work, the rms deviation increases from $0.115$~MeV to $0.385$~MeV, and the difference of the calculated and experimental value, $\Delta$BE, becomes $1.35$~MeV for the $^{94}$Ag. Importantly, the use of three-nucleon master term (3N) reduces deviation from experimental data compared two-nucleon master term (2N), matching with the findings of ab-initio theory on the need for three-body forces for mass description~\cite{O28-PhysRevLett.105.032501}. The excellent agreement between the theoretical and experimental half-life, binding energy, and energy difference of the (21$^+$) and (7$^+$) states gives confidence to the accuracy of the shell-model picture of the (21$^+$) isomer.
%This highlights the uncertainty of the AME mass extrapolation at the \textit{N=Z} line, and the need for direct mass measurements of odd-odd \textit{N=Z} nuclei in the region.
The partial half-life of the potential two-proton decay branch can be evaluated using the semi-empirical framework of Ref.~\cite{saxenaDeformationDependence2pradioactivity2022} together with the shell-model quadrupole deformation $\beta \sim 0.07$. The calculated partial half-life to the (6$^+$) ground state in $^{92}$Rh, with $Q_\mathrm{2p}$=1.4~MeV and $L$=15~$\hbar$, is approximately $10^{22}$~s. For the last energetically possible final state, (10$^+$) in $^{92}$Rh, with $Q_\mathrm{2p}$=0.140 MeV and  $L$=11~$\hbar$, an approximate partial half-life of $10^{146}$~s is obtained. This is similar to the decay of $^{54m}$Ni~\cite{giovinazzo4DimagingDriplineRadioactivity2021}, where the two-proton decay branch is energetically possible but unlikely. 

In summary, we have measured the masses of $^{93}$Pd and the long-lived excited nuclear states of $^{94}$Ag in one of the few direct mass measurements of proton-emitting species to date~\cite{kankainenMassMeasurementsVicinity2010,rauth_first_2008}. With decay $Q$ values derived with these masses, we show that the 1.9~MeV sum-peak energy reported in Ref.~\cite{Mukha2006} cannot have been caused by two-proton emission from the (21$^+$) isomer in $^{94}$Ag and that such emission is very unlikely to occur at all. We have tested two theoretical methods against our new mass data to produce a robust picture of  $^{94}$Ag and its (21$^+$) isomer. The LSSM calculations of the energy difference of the (7$^+$) and (21$^+$) states match excellently with the experimental value, and closely agree with the measured half-life of the (21$^+$) state, supporting the shell-model calculations in $N=Z$ nuclei near $^{100}$Sn. Importantly, the calculations predict the near sphericity of the (21$^+$) state. Moreover, the shell model and the $ft$ estimate clearly point towards a significantly more bound (0$^+$) ground state of $^{94}$Ag compared to AME2020~\cite{wangAME2020Atomic2021}. This large difference to AME2020, coupled with the shell-model evaluated binding energies, calls for a direct measurement of the masses of $^{94}$Ag $(0^+)$ and the neighboring odd-odd $N=Z$ nuclei. Moreover, the ab-initio VS-IMSRG calculations face challenges in describing the structure of the $^{94}$Ag states, furthermore stressing the need for new data from the region. 

Our work brings to conclusion the long-standing debate on the claimed two-proton radioactivity of the $^{94}$Ag~(21$^+$) isomer. The experimental techniques developed for and applied to this accomplishment are a significant step forward in the study of the properties of \textit{N=Z} nuclei near $^{100}$Sn. Particularly, they are a precursor for a charge radius and magnetic moment measurement of $^{94}$Ag(21$^+$) that would mark the first optical probing of a proton emitting nuclear state. 
\begin{acknowledgments}
We acknowledge the Research Council of Finland (RCF) funded this work through FIRI roadmap research infrastructures and  MR's Project No. 368322 (NUANZ). European Union’s Horizon 2020 research and innovation program grant agreement No. 771036 (ERC CoG MAIDEN) and No. 861198–LISA–H2020-MSCA-ITN-2019 are acknowledged.  D.D. Dao and F.N. acknowledge  the financial support from CNRS/IN2P3, France, via ABI-CONFI Master project. D.D.D. acknowledges the financial support from CNRS/IN2P3, France, through the budget for early-career CNRS researcher. G.K.K. acknowledges the support from the German Federal Ministry for Education and Research (BMBF) under Grant 05P21RGFN1 and 05P24RG4, from HGS-HIRe and from the German Academic Exchange Program (DAAD) under Grant 57610603  and from the UK Science and Technology Facilities Council under Grant ST/Y000293/1. A.K. acknowledges the funding from the RCF under the grants No. 354968 (HENNA), No. 369714 (NUCLEI) and No. 374065 (Centre of Excellence in Neutron-Star Physics). Z.G acknowledges RCF grant No. 354589.  B.S.H. acknowledges the Cyclotron Institute at Texas A\&M University. VS-IMSRG computations were performed with an allocation of computing resources on Fir at WestGrid and the Digital Research Alliance of Canada.
V. Virtanen and M. Reponen contributed equally to this work.
\end{acknowledgments}

\bibliographystyle{apsrev4-2}
\bibliography{ref}
\appendix
\section{Experimental setup}
\label{em: exp}
A  $^{40}$Ca beam at 8+ charge state from the K-130 cyclotron, energy degraded with molybdenum foils to match the peak of the production cross section for each isotope of interest, was impinged on a $\sim$~2.5-$\mu m$ thick enriched $^{58}$Ni target. The target foils were placed in an external rotating target wheel, located about 30 mm upstream from the hot cavity window. A 4~$\mu$m Mo  foil stack was used to reduce a 205~MeV front-of-target beam energy to $\sim$~165~MeV for $^{95}$Ag and a single 1~$\mu$m Mo foil was used to reduce the energy to 195~MeV for $^{94}$Ag. The production of $^{93}$Pd utilized a 230~MeV primary beam and a 3 $\mu$m Mo degrader to reduce the front-of-target energy to 199~MeV. The reaction products were stopped in a hot cavity catcher laser ion source~\cite{Reponen2015, reponen_evidence_2021, ge2024highprecision}, from where they effused into the graphite transfer tube. There, the atoms were selectively ionized to a single charge state using three-step laser ionization schemes tailored for silver~\cite{Reponen2015, reponen_evidence_2021, ge2024highprecision} or palladium~\cite{kron_high_2016}.

For the first resonant atomic transition~\cite{kron_high_2016} in palladium, $4\mathrm{d}^{10}\ ^1 \mathrm{S}_0 \rightarrow 4\mathrm{d}^9 \left( ^2 \mathrm{D} _{5/2} \right)5\mathrm{p}\ ^2\left[3/2 \right]^0_1 $ (Fig.~\ref{fig: setup}), the isotope shifts between the even mass stable isotopes are small, below 100 MHz~\cite{vanduijnHyperfineStructureIsotope2002a}. The wavelengths of the broadband (4 GHz in the fundamental) Ti:sapphire lasers used to drive the transitions were only optimized on less exotic isotopes. However, in $^{94}$Ag the high nuclear spins and the known large ground-state hyperfine splitting in exotic silver isotopes~\cite{reponen_evidence_2021} required matching the first step, $5\mathrm{s}\ ^2\mathrm{S}_{1/2} \rightarrow 5\mathrm{p}\ ^2\mathrm{P}_{3/2}$, laser wavelength to the expected hyperfine transitions, even if the broadband setup was insensitive to isotope or isomer shifts. As the structure of the (7$^+$) and (21$^+$) states are unknown, the estimates were based on magnetic moments calculated with the JUN45 shell-model interaction~\cite{honmaNewEffectiveInteraction2009}. Although the transitions for the tentative ($0^+$) state and the (7$^+$) state fit within the laser bandwidth of more than 15 GHz (frequency tripled from the fundamental) the laser needed to be tuned to the calculated resonance to efficiently ionize the $(21^+)$ state. Detuning the laser frequency provided an additional means of verifying that there was no laser-independent background within the time-of-flight regions corresponding to the isomeric states.

Ions were extracted into a sextupole ion guide (SPIG)~\cite{karvonen_sextupole_2008}. The ions were then electrostatically accelerated to 30~kV, mass-separated with a dipole magnet with a typical mass resolving power $M/\Delta M = 400$, and guided into the experimental area to determine their masses via multi-reflection time-of-flight measurements. The mass-over-charge $m/q$ separated ion beam was accumulated in the radiofrequency quadrupole cooler buncher \cite{NIEMINEN2001244}. The buncher potential was set to reduce the ion beam energy to approximately 100~eV, before trapping and bunching the ions in the 3~MHz, 1~kV$_\mathrm{pp}$ RF field of the Mini Buncher~\cite{virtanenMiniaturisedCoolerbuncherReduction2025}. The ion bunches, extracted from the buncher with 2~keV energy and with approximately 50~ns temporal spread, entered the MR-ToF-MS. The number of revolutions within the MR-ToF-MS\cite{virtanenHighresolutionMultireflectionTimeofflight2025} was selected to minimize the background rate at the expected time-of-flight region of interest. A single measurement was run for multiple hours to collect sufficient statistics on the time-of-flight peak of interest. Upon extraction, ion impacts were registered with a MagneToF detector~\cite{stresau2006new} and the arrival times of these signals were recorded using a FAST COMTEC MCS6A time-to-digital converter~\cite{virtanenHighresolutionMultireflectionTimeofflight2025}.

%trapped between electrostatic mirror electrodes~\cite{virtanenHighresolutionMultireflectionTimeofflight2025} by reducing their energy with a pulsed drift tube. After the different species were separated in time, they were extracted by pulsing the drift tube. Ion impacts were registered with a MagneToF detector~\cite{stresau2006new} and the times of these signals were recorded using a FAST COMTEC MCS6A time-to-digital converter (TDC). The TDC also recorded the ion release times from the cooler buncher, allowing the ion time-of flight to be reconstructed
 
The time-of-flights were determined from drift-corrected time-of-flight spectra by performing an unbinned extended maximum-likelihood fit \cite{BARLOW1990496}. The fit used a peak shape extracted from a clean reference time-of-flight spectrum by using a logarithmic natural cubic spline method, similar to Ref.~\cite{D0AN00325E}. The reference spectra were produced with an offline ion source immediately following the on-line run under the same experimental conditions. To reduce the effects of ion-ion interactions~\cite{bolotskikh_coulomb_2011} on the time-of-flight, data containing only four or fewer detected ions per bunch were selected. Time-of-flights that were seen to be strongly affected by the number of ions per bunch were corrected by a count-rate class analysis~\cite{kellerbauer_direct_2003}. This was done by extrapolating the ion's time-of-flight to 1 ion per bunch, accounting for the estimated detector efficiency of 0.7(2). Gradual time-of-flight drifts were compensated by periodically summing blocks of data and fitting the ToF of a single reference peak with a known mass. The drift-corrected time-of-flights of the ions of interest were then found by linearly interpolating between the blocks.

In the peak fitting of low count-rate ions of interest, an additional systematic uncertainty related to the selected data range for the fits and its effect on the fit result was taken into account. The resulting error terms were added to the respective covariance matrix elements of each dataset. The residual uncertainty reported in the second bracket of Table \ref{tab: massexesses} was estimated as 30~keV based on the measurement results of well-known masses of $^{95}$Ag$^+$, $^{94}$Zr$^+$ and $^{93}$Nb$^+$ .

\section{Shell-model calculations}
\label{em: sm}
%Further changes to "corrections_np_version.tex"
The large scale shell-model calculations were performed using a $^{80}$Zr  closed-shell core  and $\pi\nu$(gds) model space with  single-particle orbits 2s$_{1/2}$, 1d$_{5/2}$, 1d$_{3/2}$, 0g$_{7/2}$ and 0g$_{9/2}$ for the valence neutrons and protons~\cite{nowackiShellModelDescription2002}. The LSSM calculations used the SDGN effective interaction incorporating core excitations, which has recently been shown to perform well for nuclei in the vicinity of $^{100}$Sn~\cite{graweIsomer102SnRevisited2021} and in particular in the reproduction of the recent observation of the 2$^+$ excitation energy in $^{100}$Sn~\cite{Liu:2026:NATURE}.  
Moreover, to evaluate binding energies, we added a monopole master term from the Duflo-Zuker microscopic mass formula~\cite{DZmass94,AnatomyDZ} to the original interaction. This enables us to extract one subset of two-body matrix elements responsible for the masses, without altering the spectroscopy. Such a procedure has been used in Ref.~\cite{Caurier99-PhysRevC.59.2033}.

The most general form would consist of two and two+three body contributions for like particles and proton-neutron interactions:
\begin{equation}
\begin{aligned}
H_m = \textcolor{black}{E_0(N_0,Z_0) + \varepsilon_\nu.n + \varepsilon_\pi.z } \\ \textcolor{black}{+ V^{2N}_{\nu\nu}.\frac{n(n-1)}{2} + V^{2N}_{\pi\pi}.\frac{z(z-1)}{2} +  
 V^{2N}_{\nu\pi}.nz}  \\
 \textcolor{black}{+ V^{3N}_{\nu\nu\nu}.\frac{n(n-1)(n-2)}{6} + V^{3N}_{\pi\pi\pi}.\frac{z(z-1)(z-2)}{6}}  \\
  \textcolor{black}{+ V^{3N}_{\nu\nu\pi}.\frac{n(n-1)}{2}z + V^{3N}_{\nu\pi\pi}.n\frac{z(z-1)}{2}} 
\end{aligned}
\end{equation} where $n,z$ are the valence neutron and proton numbers, respectively.
In the present case, due to charge symmetry we keep the neutron $\varepsilon_\nu$, $V_{\nu\nu}^{2N}$/$V_{\nu\nu\nu}^{3N}$ and proton  $\varepsilon_\pi$, $V_{\pi\pi}^{2N}$/$V_{\pi\pi\pi}^{3N}$ terms identical so that we are left with 3 parameters for the quadratic form and 3 more parameters for the cubic form.  
The Coulomb energies are evaluated using the simple formula $E_C$ from ~\cite{AnatomyDZ}:
\begin{equation}
\begin{aligned}
E_C& = \frac{Z(Z-1)+0.76[Z(Z-1)^{2/3}]}{r_c}  \\
 \mathrm{with}\quad  r_c& =A^{1/3}\Bigg[1-\bigg(\frac{T}{A}\bigg)^2\Bigg]
\end{aligned}
\end{equation}
where $Z$, $T$ and $A$ are the total proton, isospin and mass numbers. 
%\begin{table}[!tb]
%\centering
%\begin{tabular}{cccc}
%          \hline\hline
%                       & {\bf} I & \textcolor{blue}{\bf II} & \textcolor{red}{\bf III}     \\[5pt]
%            \hline 
%$rmsd$( keV)  & 251 & 155 & 112  \\
%    \hline\hline
%\end{tabular}
%\label{tab:rmsd}
%\caption{rmsd in keV }
%\end{table}

Finally, the potential-energy surface for $^{94}$Ag was calculated with constrained Hartree-Fock minimization in the shell-model basis for $^{94}$Ag~\cite{daoNuclearStructureDiscrete2022}.

\section{Half-life calculation}
\label{em: hl}
Following the procedure from Refs.~\cite{A48-95,RevModPhys2005}, the partial half-life between an initial and a final nuclear state is given by
\begin{equation}
 \label{eq:ft}
 (f_A+f^\epsilon)t=\frac{(6144.48\pm3.7)~s}{(f_V/f_A)B(F)+B(GT)}.
\end{equation}
The value $(6144.48\pm3.7)~s$ is obtained from the fifteen best-known superallowed $\beta$-decays~\cite{Hardy2020}, $f_V$ and $f_A$
are the Fermi and Gamow-Teller phase-space factors, respectively~\cite{fA-WILKINSON197458}.
$f^\epsilon$ is the phase space for electron capture~\cite{EC-RevModPhys.49.77}, that is only present in $\beta^+$ decays.

$B(F)$ and $B(GT)$ are defined as
\begin{subequations}
\begin{gather}
      \label{eq:fermidef}
     B(F)=\left[\frac{\langle f \parallel \sum_{k} \  \bm{t}^{k}_{\pm} \parallel i\rangle}{\sqrt{2J_{i}+1}}\right]^2 \\ 
    \label{eq:gamowteller}
  B(GT)=\left[\left(\frac{g_A}{g_V}\right) \frac{\langle f\parallel \sum_{k} \bm{\sigma}^{k} \bm{t}^{k}_{\pm} \parallel i\rangle}{\sqrt{2J_{i}+1}}\right]^2. 
\end{gather}
\end{subequations}

Matrix elements are reduced with respect to spin only, $\pm$ refers to $\beta^\pm$ decay, $\sigma=2S$ and $(g_A/g_V)=-1.2754(13) $~\cite{gA-PhysRevD.110.030001} is the ratio of the weak interaction axial-vector and vector coupling constants.
For the (21$^+$) isomer half-life calculation, $B(GT)$ matrix elements are computed from the initial  state to 75  final states for each 20$^+$, 21$^+$ and 22$^+$ spin channel. The total half-life is then determined from the partial half-lives for each final state. 

\section{VS-IMSRG calculations}
\label{em: abinito}
Our valence-space in-medium similarity renormalization group (VS-IMSRG) calculations \cite{tsukiyama2011,hergert2016,stroberg2017,hu2022a,ye2025} start from chiral two-nucleon (NN) and three-nucleon (3N) interactions expressed in a harmonic-oscillator basis of 15 major shells with frequency $\hbar \omega=16$ MeV. An additional truncation,  $e_1+e_2+e_3 \leq (E_{\rm 3max}=24)$, is imposed for the storage of 3N matrix elements~\cite{Miyagi2022}. After transforming to the Hartree-Fock basis, we employ the Magnus formulation of the IMSRG \cite{morris2015,hergert2016,stroberg2017,miyagi2020} to construct an approximate unitary transformation that decouples a $^{80}$Zr core with the $sdg$ valence space. Through ensemble normal ordering, we effectively capture the 3N forces between valence nucleons~\cite{stroberg2017}. Finally, the resulting valence-space Hamiltonian is diagonalized using the {\tt KSHELL} code~\cite{shimizu2019},  allowing for up to $6p$-$6h$ excitations across the $N=Z=50$ shell gap.
%TC:endignore

\end{document}